%% file: ANA-SUSY-2020-17-PAPER.tex
\pdfoutput=1
\documentclass[PAPER, atlasdraft=false, texlive=2023, UKenglish, cernpreprint, texmf, orcidlogo]{atlasdoc}

\usepackage[subcaption]{atlaspackage}
\usepackage{atlasbiblatex}
 
\usepackage{atlasphysics}

\graphicspath{{logos/}{figures/}}
 
\usepackage{atlasbsm}
\usepackage{ANA-SUSY-2020-17-PAPER-defs}
\usepackage{multirow}

\AtlasTitle{Search for pair-produced higgsinos decaying via Higgs or $Z$ bosons to final states containing a pair of photons and a pair of $b$-jets with the ATLAS detector}

\AtlasAbstract{
A search is presented for the pair production of higgsinos $\tilde{\chi}$ in gauge-mediated supersymmetry models,
where the lightest neutralinos $\tilde{\chi}_1^0$ decay into a light gravitino $\tilde{G}$ in association with either a Higgs
$h$ or a $Z$ boson.
The search is performed with the ATLAS detector at the Large Hadron Collider using 139 fb$^{-1}$
of proton--proton collisions at a centre-of-mass energy of $\sqrt{s} =13\,\mathrm{TeV}$. It targets final states in which a Higgs boson decays into a photon pair, while the other Higgs or $Z$ boson decays into a $b\bar{b}$ pair,
with missing transverse momentum associated with the two gravitinos.
Search regions dependent on the amount of missing transverse momentum are defined by the requirements that the diphoton mass
should be consistent with the mass of the Higgs boson, and the $b\bar{b}$ mass with the mass of the Higgs or $Z$ boson.
The main backgrounds are estimated with data-driven methods using the sidebands of the diphoton mass distribution.
No excesses beyond Standard Model expectations are observed and higgsinos with masses up to $320\,\mathrm{GeV}$ are excluded,
assuming a branching fraction of 100\% for $\tilde{\chi}_1^0\rightarrow h\tilde{G}$. This analysis excludes higgsinos with masses of $130\,\mathrm{GeV}$ for branching fractions to $h\tilde{G}$ as low as 36\%, thus providing complementarity to previous ATLAS searches in final states with multiple leptons or multiple $b$-jets, targeting different decays of the electroweak bosons.
}

\AtlasRefCode{SUSY-2020-17}
 
\PreprintIdNumber{CERN-EP-2023-039}

\arXivId{2404.01996}
 
\HepDataRecord{2773395}
 
\AtlasJournal{Phys.\ Lett.\ B.} 
\AtlasJournalRef{\PLB 856 (2024) 138938}
\AtlasDOI{10.1016/j.physletb.2024.138938}

\hypersetup{pdftitle={ATLAS document},pdfauthor={The ATLAS Collaboration}}
 
\begin{document}
 
\maketitle

In the Minimal Supersymmetric Standard Model~\cite{Fayet:1976et,Fayet:1977yc} two neutral higgsinos, the proposed spin-half supersymmetric (SUSY)~\cite{Martin:1997ns} partners of Higgs bosons, mix together with the spin-half partners of the neutral $\mathrm{SU}(2)_L$ and $\mathrm{U}(1)_Y$ gauge bosons, forming four neutralino states $\tilde{\chi}^0_{1,2,3,4}$, while the charged higgsino mixes with the spin-half partners of the $W^\pm$ to form two pairs of chargino states $\tilde{\chi}^{\pm}_{1,2}$, where indices are ordered in increasing mass.
When the mass terms for the spin-half partners of the gauge bosons are large,
the most accessible electroweak states at colliders are the lightest,
higgsino-dominated, neutralinos $\ninoone,\ninotwo$ and charginos $\chinoonepm$.
These particles are expected to have a mass around the electroweak scale following naturalness arguments~\cite{Barbieri:1987fn,deCarlos:1993yy},
with mass splittings of $1\,\GeV$ or less.
Such mass spectra are predicted, for example, in general gauge mediation (GGM) models~\cite{Nappi:1982hm,AlvarezGaume:1981wy,Dine:1981gu,Meade:2008wd,Cheung:2007es} or in
gauge-mediated supersymmetry breaking (GMSB) models~\cite{Dine:1981gu,AlvarezGaume:1981wy,Nappi:1982hm}
in which the lightest neutralino $\ninoone$ is higgsino dominated and decays into a light gravitino (\gravino)~\cite{Kolda:1997wt}, the proposed spin-$\tfrac{3}{2}$ partner of the graviton, in association with a Higgs boson $h$ or $\Zboson$ boson. The heavier $\ninotwo$ and $\chinoonepm$ decay into the lighter $\ninoone$ in association with quarks and leptons.

These processes represent a promising supersymmetry search target at the LHC,
particularly if other SUSY particles are sufficiently heavy as to be kinematically inaccessible
or to have suppressed cross-sections.
The resulting gravitinos would not be detected by collider experiments,
but their presence could be inferred from momentum imbalance
in the direction perpendicular to the beams (\met).
The branching fractions \BFNh and \BFNZ will depend on the components of the neutralino mixing matrix. In the following, it is assumed that there are no additional decay modes of the lightest neutralino, i.e., $\BFNh + \BFNZ = 100\%$.
The subsequent decay products of the $h$ and the $\Zboson$ can be detected in a variety of different decay modes. Decays of the lightest neutralino into photons are also possible, but result in significantly different final states, which are considered in other searches, e.g. Ref.~\cite{SUSY-2018-11}.
 
\begin{figure}[t!b]
\centering
\includegraphics[width=.5\textwidth]{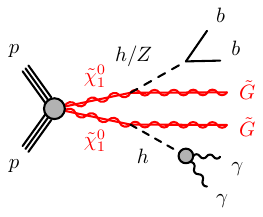}
\caption{\small  The diagram shows the topology considered by this search, the production of two neutralinos \ninoone, which decay into a gravitino \gravino and either a Higgs boson $h$ or a $Z$ boson. The search aims to select events in which a pair of photons is produced by a Higgs boson from one \ninoone, while a pair of $b$-quarks is produced by either the decay of a Higgs boson or a $Z$ boson from the other.}
\label{fig:Feynman}
\end{figure}

This paper presents a search that targets final states in which a Higgs boson decays into a photon pair, while the other Higgs boson decays into a $b\bar{b}$ pair,
with missing transverse momentum associated with the two gravitinos.
This final state also has sensitivity to mixed decays in which one neutralino decays into $h$ and one into $\Zboson$ where the \Zboson decays into a pair of $b$-quarks, as shown in Figure~\ref{fig:Feynman}. The ATLAS Collaboration has previously performed searches using events with at least three $b$-jets~\cite{ATLAS:2024tqe}, in $h\rightarrow \gamma\gamma$ events without requiring the $b$-jet decay mode of the other boson~\cite{SUSY-2018-23},
events with four or more leptons~\cite{SUSY-2018-02}, events containing two leptons, jets and missing transverse momentum~\cite{SUSY-2018-05}, and in events with two boosted bosons, missing transverse momentum and no leptons~\cite{SUSY-2018-41}.
The CMS collaboration has performed several searches in a variety of decay channels~\cite{CMS-SUS-14-002,CMS-SUS-16-044,CMS-SUS-18-007}
including $\gamma\gamma b b$ ~\cite{CMS-SUS-18-007}.
For the searches involving two Higgs bosons in the final state, decays into $\gamma\gamma b b$ combine the advantages of excellent mass resolution in $h\rightarrow \gamma\gamma$
with the large Higgs boson branching fraction to $b\bar{b}$.
 
The analysis is based on data collected by the ATLAS detector~\cite{PERF-2007-01} using proton--proton ($pp$) collisions delivered by the LHC~\cite{Evans:2008zzb} between 2015 and 2018, at a centre-of-mass energy of $\sqrt{s}=13\,\TeV$ with a bunch-crossing interval of 25\,ns. The average number of $pp$ interactions per bunch crossing ranged from about 13 in 2015 to about 39 in 2018.
After applying trigger and data-quality requirements~\cite{DAPR-2018-01} the integrated luminosity of the data sample,
measured with the LUCID-2 detector~\cite{Avoni:2018iuv} using the methods described in Ref.~\cite{ATLAS-CONF-2019-021}, corresponds to $139.0 \pm 2.4\,\textrm{fb}^{-1}$.

The ATLAS detector at the LHC is a multipurpose particle detector
with a forward--backward symmetric cylindrical geometry and a near \(4\pi\) coverage in
solid angle.\footnote{ATLAS uses a right-handed coordinate system with its origin at the nominal interaction point (IP)
in the centre of the detector and the \(z\)-axis along the beam pipe.
The \(x\)-axis points from the IP to the centre of the LHC ring,
and the \(y\)-axis points upwards.
Polar coordinates \((r,\phi)\) are used in the transverse plane,
\(\phi\) being the azimuthal angle around the \(z\)-axis.
The pseudorapidity is defined in terms of the polar angle \(\theta\) as \(\eta = -\ln \tan(\theta/2)\).
Angular distance is measured in units of \(\Delta R \equiv \sqrt{(\Delta\eta)^{2} + (\Delta\phi)^{2}}\).}
It consists of an inner tracking detector (ID) surrounded by a thin superconducting solenoid
providing a 2~T axial magnetic field, electromagnetic and hadron calorimeters, and a muon spectrometer (MS).
The inner tracking detector covers the pseudorapidity range \(|\eta| < 2.5\).
It consists of silicon pixel, silicon microstrip, and transition radiation tracking detectors.
Lead/liquid-argon (LAr) sampling calorimeters provide electromagnetic (EM) energy measurements
with high granularity.
A steel/scintillator-tile hadron calorimeter covers the central pseudorapidity range \(|\eta| < 1.7\).
The endcap and forward regions are instrumented with LAr calorimeters
for both the EM and hadronic energy measurements up to \(|\eta| = 4.9\).
The muon spectrometer surrounds the calorimeters and is based on
three large superconducting air-core toroidal magnets with eight coils each.
The field integral of the toroids ranges between \num{2.0} and $6.0\,\mathrm{Tm}$
across most of the detector.
The muon spectrometer includes a system of precision tracking chambers and fast detectors for triggering.
A two-level trigger system is used to select events.
The first-level trigger is implemented in hardware and uses a subset of the detector information
to accept events at a rate below $100\,\mathrm{kHz}$.
This is followed by a software-based trigger that
reduces the accepted event rate to $1\,\mathrm{kHz}$ on average,
depending on the data-taking conditions.
An extensive software suite~\cite{ATL-SOFT-PUB-2021-001} is used in data simulation, in the reconstruction
and analysis of real and simulated data, in detector operations, and in the trigger and data acquisition
systems of the experiment.
 
Monte Carlo (MC) simulations are used to design and optimise the search strategy,
for calculating the expected yields of signal events,
and as part of the background estimation process.
The simulated backgrounds include prompt diphoton production ($\gamma\gamma$) as well as resonant Higgs production via \ttH, vector boson fusion (VBF), \VH, gluon--gluon fusion (ggF), \tHq, \tHW and \bbH.
The main details regarding the generator settings for such processes are summarised in Table \ref{tab:BkgMC}.
In addition, for the prompt diphoton production the next-to-leading-order (NLO) matrix elements for up to one parton, and leading-order (LO) matrix elements for up to three partons were calculated with the Comix~\cite{Gleisberg:2008fv} and
\OPENLOOPS~\cite{Buccioni:2019sur,Cascioli:2011va,Denner:2016kdg} libraries. They were matched
to the \SHERPA parton shower~\cite{Schumann:2007mg} using the \MEPSatNLO
prescription~\cite{Hoeche:2011fd,Hoeche:2012yf,Catani:2001cc,Hoeche:2009rj}
with a dynamic merging cut~\cite{Siegert:2016bre} of 10~\GeV.
Photons were required to be isolated according to a smooth isolation-cone criterion~\cite{Frixione:1998jh}.
The ggF process is simulated with next-to-next-to-leading-order (NNLO) accuracy for arbitrary inclusive \(gg\to h\) observables by reweighting the Higgs boson
rapidity spectrum in \textsc{Hj}-\MINLO~\cite{Hamilton:2012np,Campbell:2012am,Hamilton:2012rf} to that of HNNLO~\cite{Catani:2007vq}.
In the \tHW{} simulated sample, the overlap with the \ttH{} production was removed using the diagram removal scheme~\cite{Frixione:2008yi,Demartin:2016axk}.
The decays of bottom and charm hadrons were simulated using the \EVTGEN[1.6.0] program~\cite{Lange:2001uf} for all Higgs boson samples. The normalisation of these samples accounts for the decay branching ratio calculated with \textsc{HDECAY}~\cite{Djouadi:1997yw,Spira:1997dg,Djouadi:2006bz} and \PROPHECY~\cite{Bredenstein:2006ha,Bredenstein:2006rh,Bredenstein:2006nk}.
All the MC samples including Higgs bosons are simulated assuming  $\mH=125~\GeV$.

In the following, the backgrounds arising from resonant Higgs processes will be categorised as \ttH and $h$ (other), summarising events from VBF, \VH, ggF, \tHq, \tHW and \bbH production, which account for only a minor contribution to the total background in the signal regions (SR).

\begin{table}[t!b]\renewcommand\arraystretch{1.2}
\centering
\caption{\small Background MC simulated samples used in this analysis. The matrix element and parton shower generators, underlying-event tune, PDF set and cross-section order are provided.}
\resizebox{\textwidth}{!}{
\begin{tabular}{lccccc}
\toprule
\textbf{Process} & \textbf{Generator} & \textbf{Parton shower} & \textbf{Tune} & \textbf{PDF} & \textbf{Cross-section order} \\
& & \textbf{\& hadronization} & & &  \\
\midrule
$\gamma\gamma$ & \SHERPA[2.2.4]~\cite{Bothmann:2019yzt} & \SHERPA parton shower~\cite{Schumann:2007mg} & \SHERPA standard & \NNPDF[3.0nnlo]~\cite{Ball:2014uwa} & data-driven \\
\ttH & \POWHEGBOX[v2]~\cite{Hamilton:2013fea,Hamilton:2015nsa,Alioli:2010xd,Nason:2004rx,Frixione:2007vw} & \PYTHIA[8.2]~\cite{Sjostrand:2014zea} & A14~\cite{ATL-PHYS-PUB-2014-021} & \NNPDF[2.3lo]~\cite{Ball:2014uwa} & NLO~\cite{deFlorian:2016spz} \\
VBF & \POWHEGBOX[v2] & \PYTHIA[8.2] & \AZNLO~\cite{STDM-2012-23} & \PDFforLHC[15nlo]~\cite{Butterworth:2015oua} & NNLO+NLO~\cite{Ciccolini:2007jr,Ciccolini:2007ec,Bolzoni:2010xr} \\
\VH & \POWHEGBOX[v2] & \PYTHIA[8.2] & \AZNLO & \PDFforLHC[15nlo] & NLO+NNLL~\cite{Ciccolini:2003jy,Brein:2003wg,Brein:2011vx,Altenkamp:2012sx,Denner:2014cla,Brein:2012ne,Harlander:2014wda}\\
ggF & \POWHEGBOX[v2] & \PYTHIA[8.2] & \AZNLO & \PDFforLHC[15nnlo] & NNLO~\cite{Hamilton:2012np,Campbell:2012am,Hamilton:2012rf,Catani:2007vq} \\
\tHq & \MGNLO[2.6.0]~\cite{Alwall:2014hca} & \PYTHIA[8.2] & A14 & \NNPDF[2.3lo] & NLO~\cite{deFlorian:2016spz} \\
\tHW{} & \MGNLO[2.6.2] & \PYTHIA[8.2] & A14 & \NNPDF[2.3lo] & NLO~\cite{deFlorian:2016spz} \\
\bbH & \MGNLO[2.2.3] & \PYTHIA[8.1]~\cite{Sjostrand:2007gs} & A14 & \NNPDF[2.3lo] & NLO~\cite{deFlorian:2016spz} \\
\bottomrule
\end{tabular}
}
\label{tab:BkgMC}
\end{table}
 
The signal MC simulation includes the production of pairs of higgsinos through initial production of
$\chinoonep\chinoonem$, $\chinoonepm\ninoone$, $\chinoonepm\ninotwo$,
and $\ninoone\ninotwo$.
The $\chinoonepm$ and $\ninotwo$ are assumed to have $m(\chinoonepm) - m(\ninoone) = 1\,\GeV$ and $m(\ninotwo) - m(\ninoone) = 1\,\GeV$ and to decay via $\chinoonepm\rightarrow ff'\ninoone$ and $\ninotwo\rightarrow ff' \ninoone$, where $f$, $f'$ are soft fermions that remain undetected.
The mass of the gravitino was set to 1\,\MeV, and the decays
$\ninoone\rightarrow h\gravino$ and $\ninoone\rightarrow\Zboson\gravino$ were similarly assumed to be prompt.
Events were generated from LO matrix elements with up to two extra partons, using the \MGNLO[2.9.3]~\cite{Alwall:2014hca}
generator interfaced to \PYTHIA[8.245]. Events were matched using the CKKW-L prescription~\cite{Lonnblad:2001iq}, with a matching scale set to one quarter of the mass of the pair-produced SUSY particles.
Signal cross-sections were calculated to NLO in the strong coupling constant, adding the resummation of soft gluon emission at next-to-leading-logarithm accuracy (NLO+NLL)~\cite{Beenakker:1996ch,Kulesza:2008jb,Kulesza:2009kq,Beenakker:2009ha,Beenakker:2011fu,Fuks:2012qx,Fuks:2013vua,Fuks:2013lya}. The cross-section for $m(\ninoone)=150\,\GeV$ amounts to $3.83\pm0.16\,\si{\pb}$.
 
The generation of all simulated signal samples includes the effect of multiple
interactions in the same and neighbouring bunch
crossings (\pileup), modelled by overlaying the simulated hard-scattering event with
inelastic \(pp\) events generated with \PYTHIA[8.186]
using the \NNPDF[2.3lo] set of parton distribution functions~\cite{Ball:2014uwa} and the
A3 set of tuned parameters~\cite{ATL-PHYS-PUB-2016-017}.
 
The various background processes contributing to the signal regions are
categorised into two classes:
resonant and non-resonant backgrounds. The resonant backgrounds are those that contain Higgs boson decays into $\gamma\gamma$,
and so are expected to have a peak in \myy{} near \mH.
They are subdominant, and are determined from MC simulations.
 
For this analysis, photons are reconstructed~\cite{EGAM-2018-01} with pseudorapidity $|\eta|<2.37$, while vetoing those photons in the EM calorimeter transition region $1.37<|\eta|<1.52$ between the calorimeter barrel and endcap, and are required to have $\pT>\SI{25}{\GeV}$. Signal photons are further required to satisfy \tight{} selection and isolation requirements~\cite{EGAM-2018-01}. Photons with a looser (\looseprime{3}~\cite{EGAM-2018-01}) identification and without any isolation requirement, defined as baseline photons, are used to aid the data-driven estimation of some of the backgrounds, as described below.
 
Jets are reconstructed using the particle-flow algorithm, which combines both calorimeter and tracking information~\cite{PERF-2015-09}.
The \antikt algorithm~\cite{Cacciari:2008gp,Fastjet} with a radius parameter of \(R=0.4\) is used, with a four-momentum recombination scheme.
From the selected jets, $b$-jets are identified using the DL1r tagger at the 77\% efficiency working point, as determined from simulated $t\bar{t}$ events~\cite{FTAG-2019-07}.
 
Electrons are reconstructed from isolated EM calorimeter energy deposits matched to ID tracks and are required to have $|\eta|<2.47$, a transverse momentum $\pT>\SI{4.5}{\GeV}$, and to satisfy the `LooseAndBLayer' requirement defined in Ref.~\cite{EGAM-2018-01}, which is based on a likelihood using measurements of shower shapes in the calorimeter and track properties in the ID as input variables.
Muons are reconstructed in the region $|\eta|<2.7$ from MS tracks matching ID tracks.
They are required to have $\pT>\SI{4}{\GeV}$ and satisfy the `Medium' identification requirements~\cite{MUON-2018-03}, based on the number of hits in the different ID and MS subsystems, and on the significance of the charge to momentum ratio, $q/p$.
 
To avoid reconstruction ambiguities and double counting of analysis objects, electrons, muons and jets are removed from events if they are within $\Delta R=0.4$ of any photon passing \loose{} identification criteria~\cite{EGAM-2018-01}. While this selection favours photons originating from electron bremsstrahlung over the parent electron itself, such photon candidates do not satisfy the photon isolation requirements and therefore do not enter the signal regions. Electrons and muons within $\Delta R=0.4$ of a jet are also removed.
 
The missing transverse momentum, \met{}, is constructed~\cite{ATLAS-CONF-2018-023} from the negative sum of all transverse momenta $-\sum_i (\vec{p}_{\text{T}})^i$ of all reconstructed objects $i$ in the event. These consist of photons, jets, muons and electrons, as well as a track soft-term that includes the transverse momenta of all tracks associated with the primary vertex\footnote{The primary vertex is defined to be that with the largest $\sum \pT^2$ of associated tracks.} that are not reconstructed as one of the particles listed.

Candidate events were selected using a combination of diphoton triggers~\cite{TRIG-2018-05,TRIG-2019-04}, and are required to have exactly two signal photons.
To ensure good trigger and reconstruction efficiency the highest-\pT (leading) photon is required to have $\pT>35\,\GeV$, and the sub-leading photon to have $\pT>25\,\GeV$.
Additionally, the diphoton invariant mass is required to lie in the range of $95\,\GeV<\myy<\,160\,\GeV$, and each photon is required to satisfy $\pT/\myy>0.2$.
Events are required to have exactly two $b$-tagged jets, and any event containing one or more
reconstructed electrons or muons is rejected.
 
Three different signal regions are then defined (Table~\ref{tab:SRdef}),
to gain sensitivity to different $\ninoone$ mass hypotheses and to the different decay modes.
The three regions are designed to be non-overlapping,
allowing a statistical combination that further enhances sensitivity.
The SR selections are optimised based on studies of simplified GMSB models with varying \ninoone mass, and of differing values of the branching fraction \BFNh.
 
\begin{table}[t!b]\renewcommand\arraystretch{1.2}
\centering
\caption{\small Definition of the three different signal region selections, where SR1h is aimed at final states containing two Higgs bosons, SR1Z is aimed at final states containing one Higgs boson and one $Z$ boson, and SR2 is aimed at higher-mass neutralinos decaying through either channel.}
\begin{tabular}{l|c|c|c}
\toprule
\textbf{Requirement} & \textbf{SR1h} & \textbf{SR1Z} & \textbf{SR2} \\
\midrule
$\myy$ & \multicolumn{3}{c}{$\in(120,130)\,\GeV$} \\\midrule
\MET  & \multicolumn{2}{c|}{$\leq100\,\GeV$} & $>100\,\GeV$ \\\midrule
$\mbb$ & $\in(100,140)\,\GeV$ & $\in(60,100)\,\GeV$ & $\in(35,145)\,\GeV$ \\\midrule
$p_{\mathrm{T}}^{\gamma\gamma}$  & \multicolumn{2}{c|}{$\geq 90\,\GeV$} & -- \\\midrule
$p_{\mathrm{T}}^{\gamma\gamma}/\myy$  & \multicolumn{2}{c|}{$\geq 0.4$} & $\geq 0.2$ \\
\bottomrule
\end{tabular}
\label{tab:SRdef}
\end{table}
 
Each of the three SRs requires a diphoton invariant mass \myy{} within 5~\GeV~of the observed mass of the Higgs boson of 125\,\GeV~\cite{HIGG-2022-20, CMS-HIG-19-004}.
The signal regions differ in their requirements on the invariant mass \mbb{} of the $b$-tagged jets,
and in the requirement on the missing transverse momentum.
SR1h targets those events that contain two Higgs bosons,
and so has an $\mbb$ requirement that is also broadly consistent with \mH.
A second region, SR1Z, is designed to select events where $\mbb$ is instead
consistent with $\mZ$,
thus gaining sensitivity to events in which one $\ninoone$ decays into
$h\gravino$ and the other into $\Zboson\gravino$.
These two regions, SR1h and SR1Z,
both target the lower end of the allowed $\ninoone$ mass range, a region in which $\MET$ is relatively small with $\MET<100\,\GeV$. Both regions include an additional selection on the transverse momentum of the diphoton system $p_{\mathrm{T}}^{\gamma\gamma}$, as well as on $p_{\mathrm{T}}^{\gamma\gamma}/\myy$, to improve the signal-to-background ratio.
 
SR2 targets higher-mass $\ninoone$ decays, which consequently demands high missing transverse momentum, $\MET>100\,\GeV$.
The expected yield for these higher-mass models is small,
so to enhance sensitivity SR2 has a rather loose selection on \mbb,
which can admit both Higgs and $\Zboson$ boson decays. No selection on $p_{\mathrm{T}}^{\gamma\gamma}$ is included to retain more signal events.
 
The dominant backgrounds are from non-resonant
events, those for which no peak at $\myy\approx\mH$ is expected.
They include prompt diphoton events and also
those with fewer than two prompt photons
that can contribute when other detected particles -- most often those associated with jets
-- are falsely identified as photons.
These prompt and fake photon non-resonant
contributions to the signal regions are modelled using data-driven techniques,
since there is a variety of different ways in which mismeasurements can contribute,
not all of which are expected to be well-modelled in MC simulations.
 
The data-driven method used to determine the non-resonant backgrounds is known as
the `\twobytwod{} sideband method', a technique developed in the context of
$h\rightarrow \gamma\gamma$ measurements~\cite{STDM-2011-05},
and based on similar techniques used at the D0~\cite{D0:2010xdg} and CDF~\cite{CDF:2011nuh} experiments.
The method makes extensive supporting measurements
in data to improve the accuracy of the predictions from MC simulations.
An outline of the method is given below, while further details may be found in Ref.~\cite{STDM-2011-05}.
 
Control regions are defined as outlined below, selecting events in the diphoton sidebands ($\myy < 120\,\GeV$ and $\myy > 130\,\GeV$)
using baseline photons. The looser requirements provide a large yield of prompt and fake photons, while
remaining unbiased by the photon identification requirements of the triggers.
These events are then partitioned into categories, based on the
photon \emph{identification} status (tight or not-tight)
and the photon \emph{isolation} status (isolated or non-isolated).
The partitioning is performed separately for the leading
and sub-leading photon candidates,
thus providing 16 different observable categories of events.
 
The expected yields of $\gamma\gamma$, $\gamma$--jet ($\gamma j$), jet--$\gamma$ ($j\gamma$) and jet--jet ($jj$) events in each region can be expressed separately using the yields obtained by selecting baseline photons, the prompt photon identification and isolation efficiencies, the fake-photon mis-identification and mis-isolation probabilities (fake-factors) and the relevant correlations between the fake-factors. It is assumed, based on previous studies~\cite{STDM-2011-05},
that any photon--photon efficiency correlations,
photon--jet efficiency correlations or fake-photon mis-identification correlations are negligible.
 
A simultaneous likelihood fit constrained by the yields in each of the 16 observable categories is then performed to determine the parameters above, where the prompt photon identification and isolation efficiencies are determined using the diphoton MC sample.
From the fitted parameters the fractions of $\gamma\gamma$, $\gamma j$, $j\gamma$ and $jj$ events in the control regions are calculated and applied to the corresponding signal and validation regions (VR) defined below.
The key assumption, that the fraction of the events of each of these four types
is the same between the control region and the corresponding signal region,
is justified by the smooth behaviour in $\myy$ of the non-resonant
backgrounds across and between the sidebands, and by the fact that the photon identification and isolation efficiencies do not depend strongly on $\myy$~\cite{HIGG-2016-21}.
 
Since the background composition is expected to change in regions with different \met{} requirements, backgrounds are estimated separately for the low-\met{} signal regions SR1h and SR1Z, and for the high-\met{} signal region SR2. All control regions select events in the \myy~sidebands ($\myy < 120\,\GeV$ and $\myy > 130\,\GeV$). Additionally, the low-\met{} control region CR1 selects events with $\met{}\leq100\,\GeV$ and $60\,\GeV < \mbb < 140\,\GeV$, while the high-\met{} control region CR2 selects events with $\met{}>100\,\GeV$ and $35\,\GeV < \mbb < 145\,\GeV$. The \met{} and $\mbb$ selections applied in the control regions therefore match the corresponding signal region(s). To overcome the limited number of events in the high-\met{} signal region, additional supporting measurements in the $\myy$ sidebands in regions defined by inverting the \met{} and/or the $b$-jet multiplicity requirements, are also used in conjunction with those closer to SR2, and the final background prediction is obtained by extrapolating from such regions with the aid of the ABCD method~\cite{Kasieczka:2020pil}.
 
\begin{figure*}[t!b]
\centering
\begin{subfigure}[b]{0.49\textwidth}
\centering
\includegraphics[width=\textwidth]{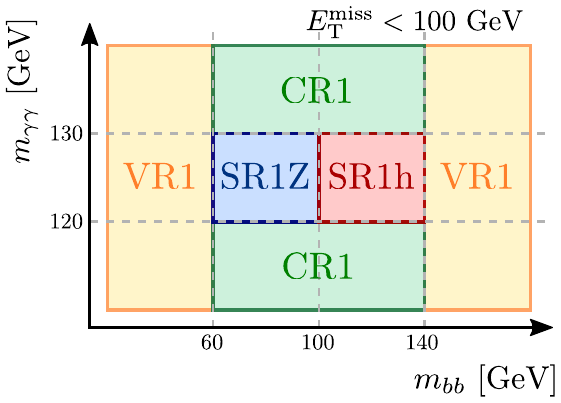}
\caption[]{}
\end{subfigure}
\begin{subfigure}[b]{0.49\textwidth}
\centering
\includegraphics[width=\textwidth]{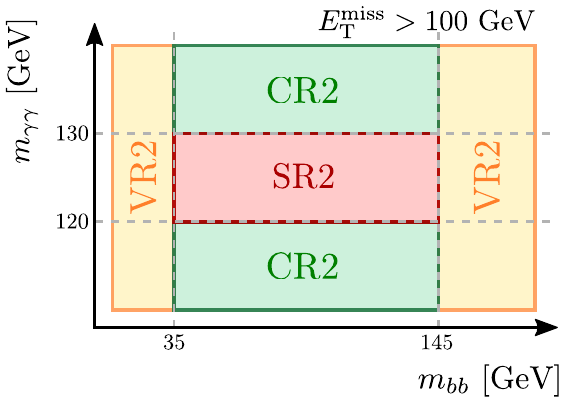}
\caption[]{}
\end{subfigure}
\caption{\small Schematic overview of the event categorisation. Events with (a) low-\MET and (b) high-\MET are treated separately, while the invariant masses of the photons $\myy$ and $b$-jets $\mbb$ are used to categorise events. The resulting control regions (CR) are used to estimate backgrounds. This estimation is applied to both the validation regions (VR) and the signal regions (SR).}
\label{fig:regions}
\end{figure*}

To test the accuracy of the background prediction, two VRs are defined.
These regions have similar kinematic selections to the signal regions,
except that they remove the requirement on $\myy$ while they
require that $\mbb$ be outside the selection band of the
corresponding signal region, resulting in negligible signal contamination ($\lesssim3\,\%$).
The two VRs are designed to reflect the same \MET spectra of the SRs, with VR1 requiring $\MET<100\,\GeV$ and VR2 requiring $\MET>100\,\GeV$.
The background estimation method described above is applied to the VRs in the same manner as for the SRs, where correction factors for the predicted numbers of $\gamma\gamma$, $\gamma j$, $j\gamma$ and $jj$ events are determined in the respective CR and applied in both the VR and the SR\@. A schematic overview of the signal, control- and validation-region strategy is shown in Figure~\ref{fig:regions}.
 
\begin{figure*}[t!b]
\centering
\begin{subfigure}[b]{0.49\textwidth}
\centering
\includegraphics[width=\textwidth]{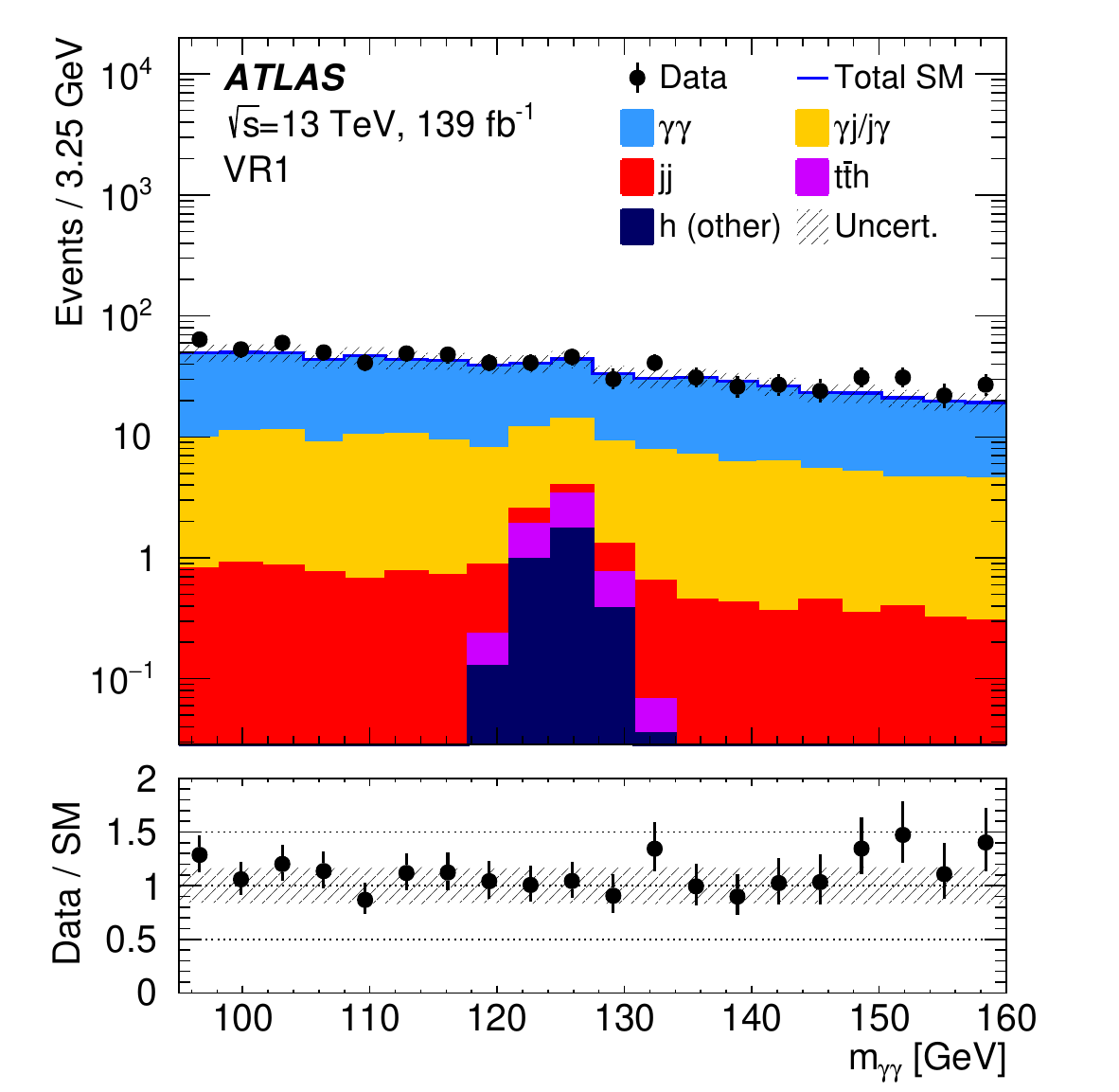}
\caption[]{}
\end{subfigure}
\begin{subfigure}[b]{0.49\textwidth}
\centering
\includegraphics[width=\textwidth]{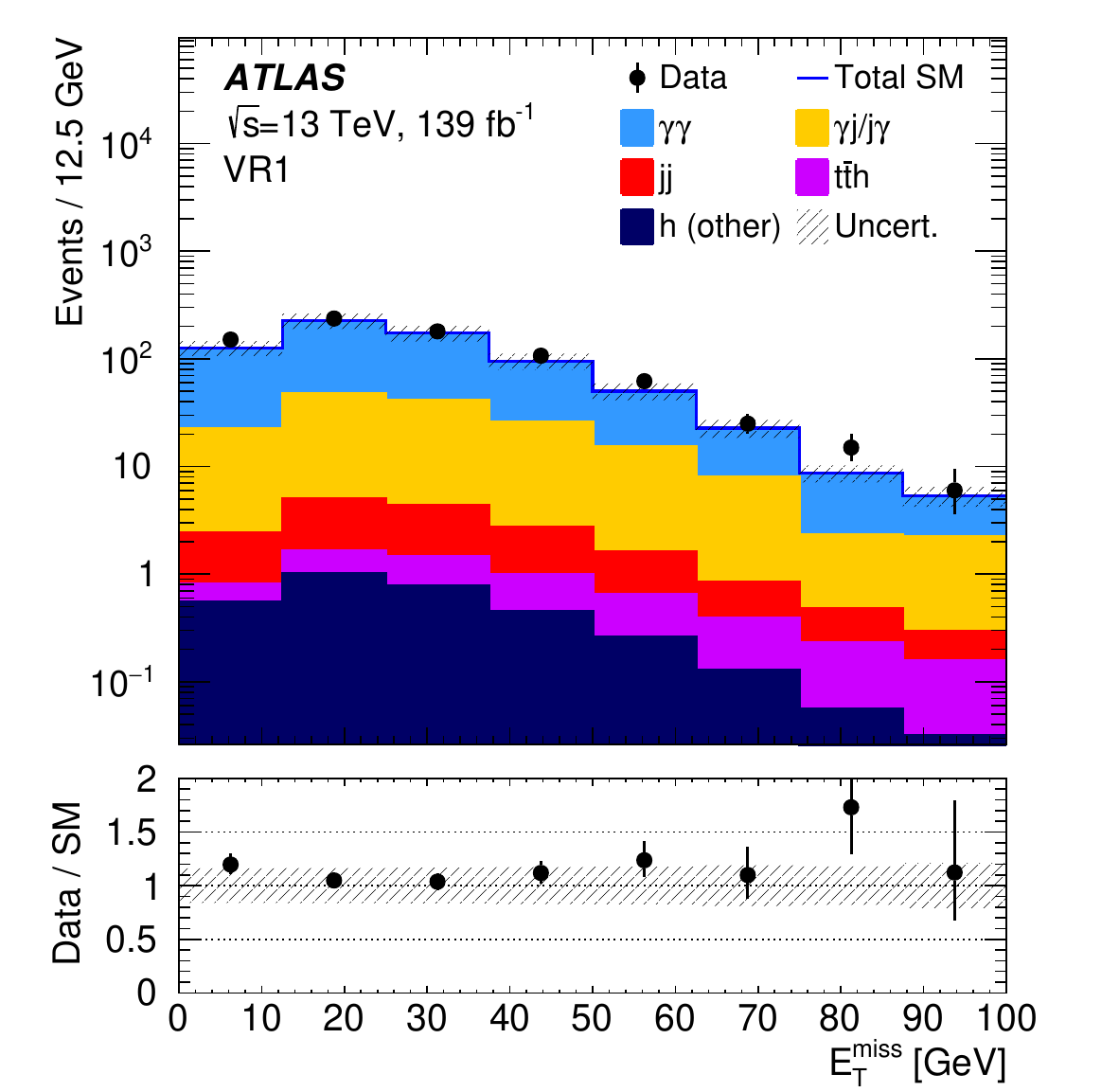}
\caption[]{}
\end{subfigure}
\begin{subfigure}[b]{0.49\textwidth}
\centering
\includegraphics[width=\textwidth]{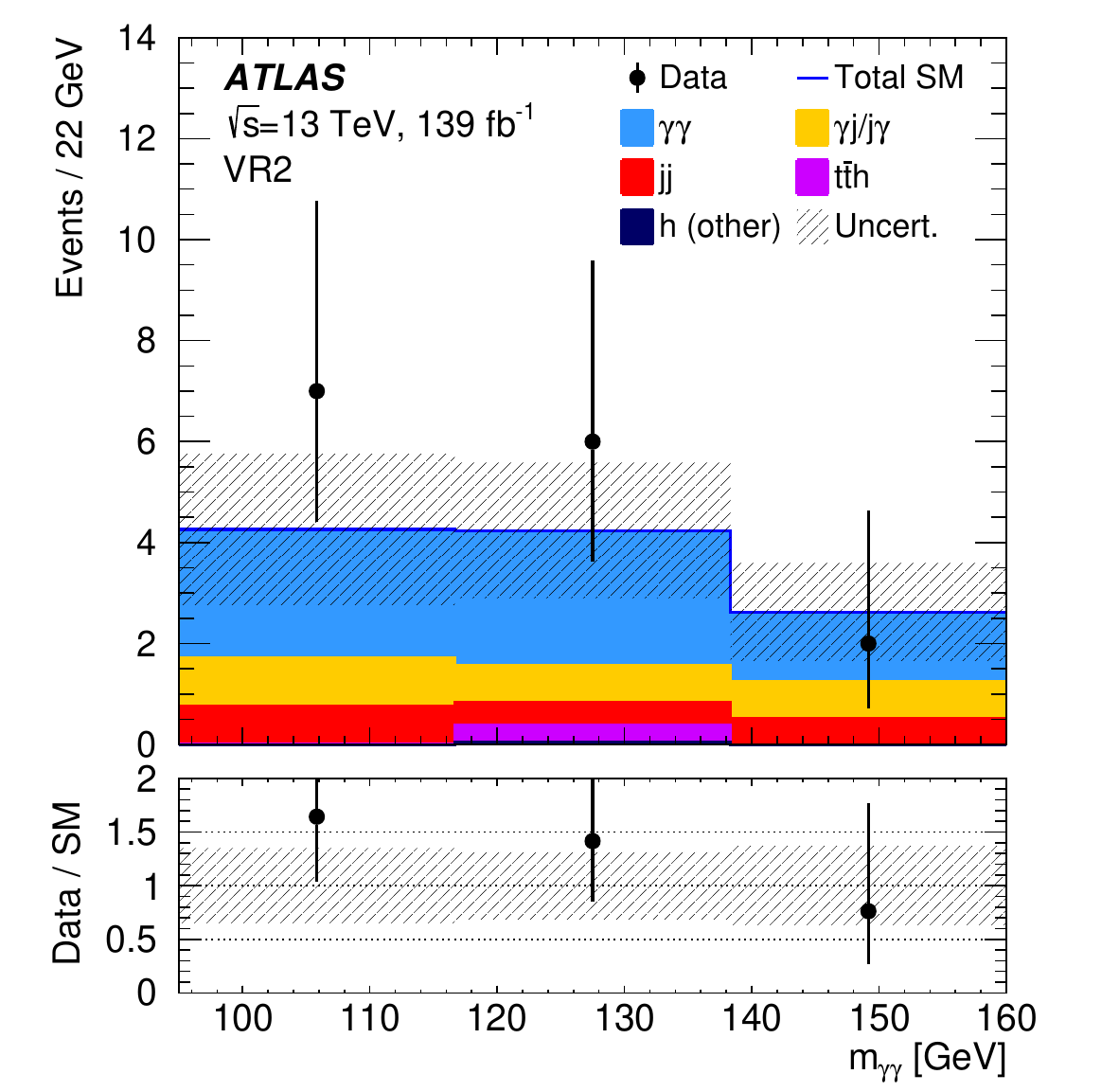}
\caption[]{}
\end{subfigure}
\begin{subfigure}[b]{0.49\textwidth}
\centering
\includegraphics[width=\textwidth]{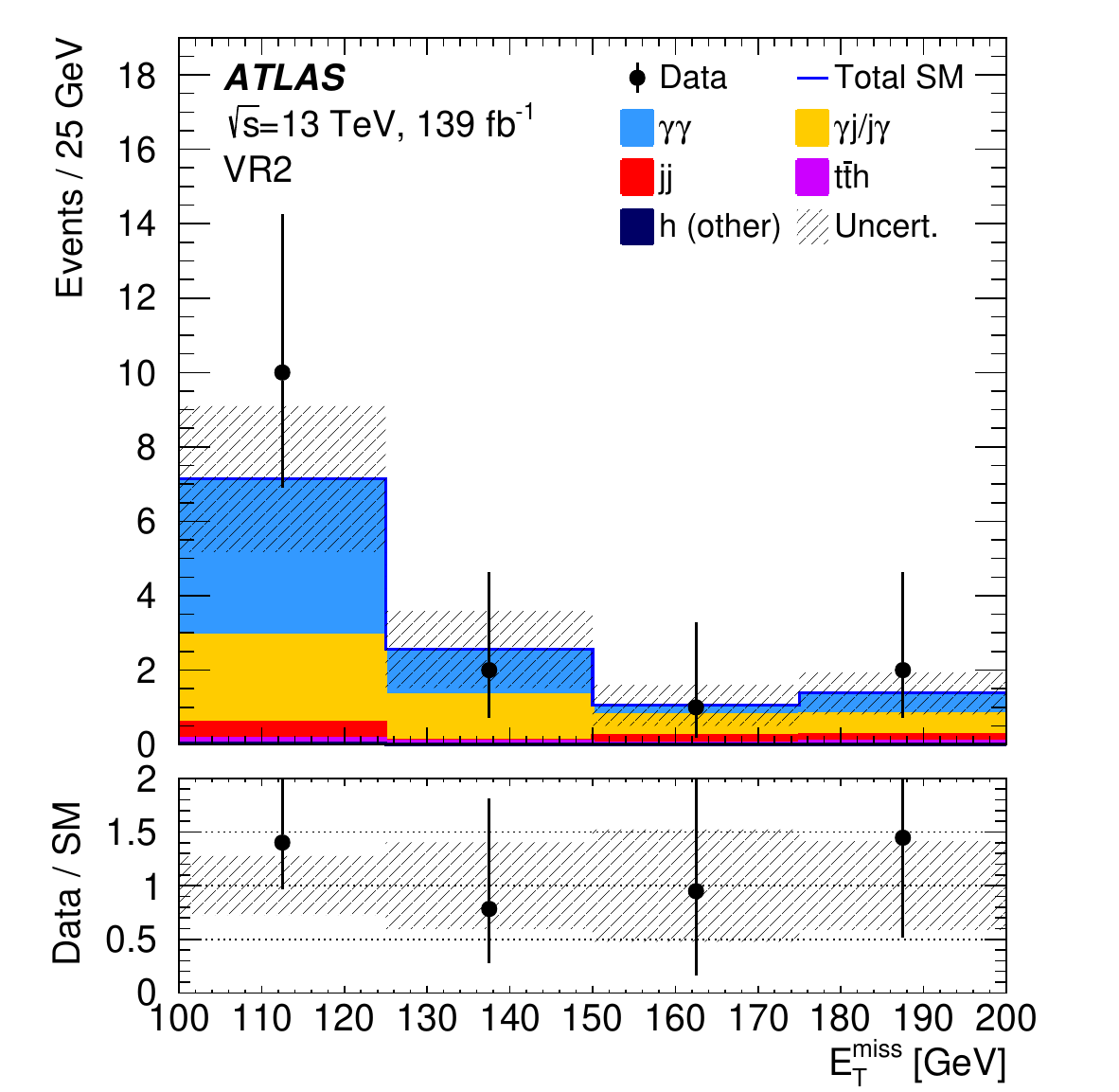}
\caption[]{}
\end{subfigure}
\caption{\small Distribution of the diphoton invariant mass in both validation regions: (a) VR1 and (c) VR2 as well as the missing transverse momentum in both validation regions: (b) VR1 and (d) VR2. The solid histograms are stacked to show the SM expectations after the \twobytwod{} background estimation technique is applied. Background and signal predictions are normalised to the luminosity. The background category `$h$ (other)' includes events originating from VBF, \VH, ggF, \tHq, \tHW and \bbH, all subdominant in this signature. Statistical and systematic uncertainties are indicated by the shaded area. The lower panel of each plot shows the ratio of the data to the SM prediction for the respective bin.
The first and last bins include the underflows and overflows respectively.}
\label{fig:VR1}
\end{figure*}

The overall yields and the shapes of the backgrounds in both VRs are found to be well predicted.
Distributions of two of the most important kinematic variables and the corresponding agreement between data and the Standard Model (SM) prediction are illustrated in Figure~\ref{fig:VR1}.
 
\FloatBarrier

Systematic uncertainties are calculated for three classes of sources:
from data-driven background methods,
from the modelling of the detector response and
from theoretical calculations of different processes that are used to normalise the corresponding MC predictions.
The total size of the systematic uncertainty in the prediction for the background yields ranges from $15\%$ to $39\%$,
depending on the signal region.
The dominant sources of systematic uncertainty are from the weights
calculated by the \twobytwod{} data-driven background method,
and from the corresponding statistical uncertainty of the associated control regions.
Systematic uncertainties associated with the detector response are found to be negligible.
The theoretical uncertainties are found to be at the level of
$3\%$ to $6\%$ depending on the SR\@.
The statistical uncertainties are in the range of $40\%$ to $77\%$,
depending on the signal region, and so are the dominant source of uncertainty
for the available data sample.
 
\begin{figure*}[p!]
\centering
\begin{subfigure}[b]{0.49\textwidth}
\centering
\includegraphics[width=\textwidth]{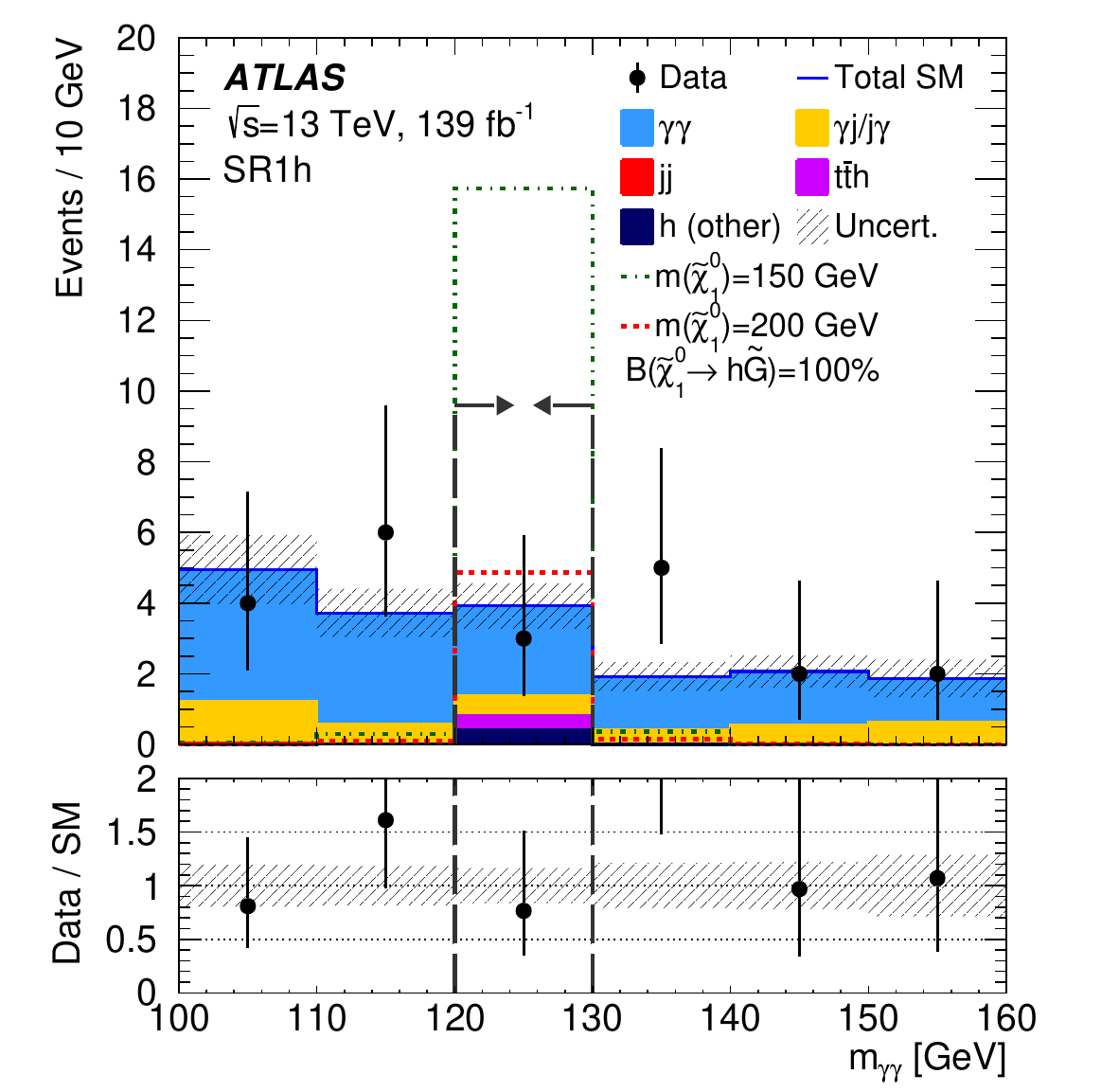}
\caption[]{}
\end{subfigure}
\hfill
\begin{subfigure}[b]{0.49\textwidth}
\centering
\includegraphics[width=\textwidth]{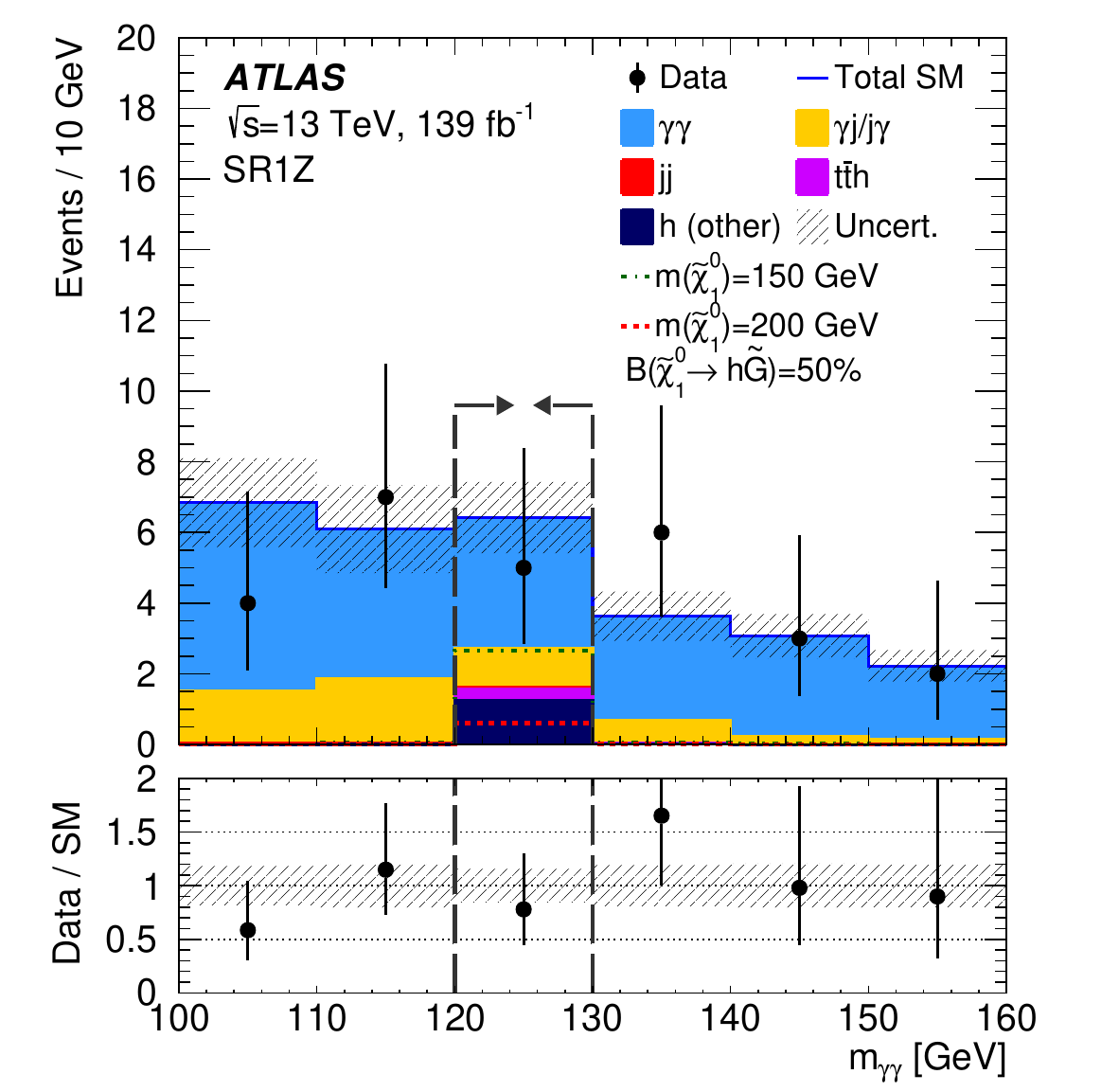}
\caption[]{}
\end{subfigure}
\begin{subfigure}[b]{0.49\textwidth}
\centering
\includegraphics[width=\textwidth]{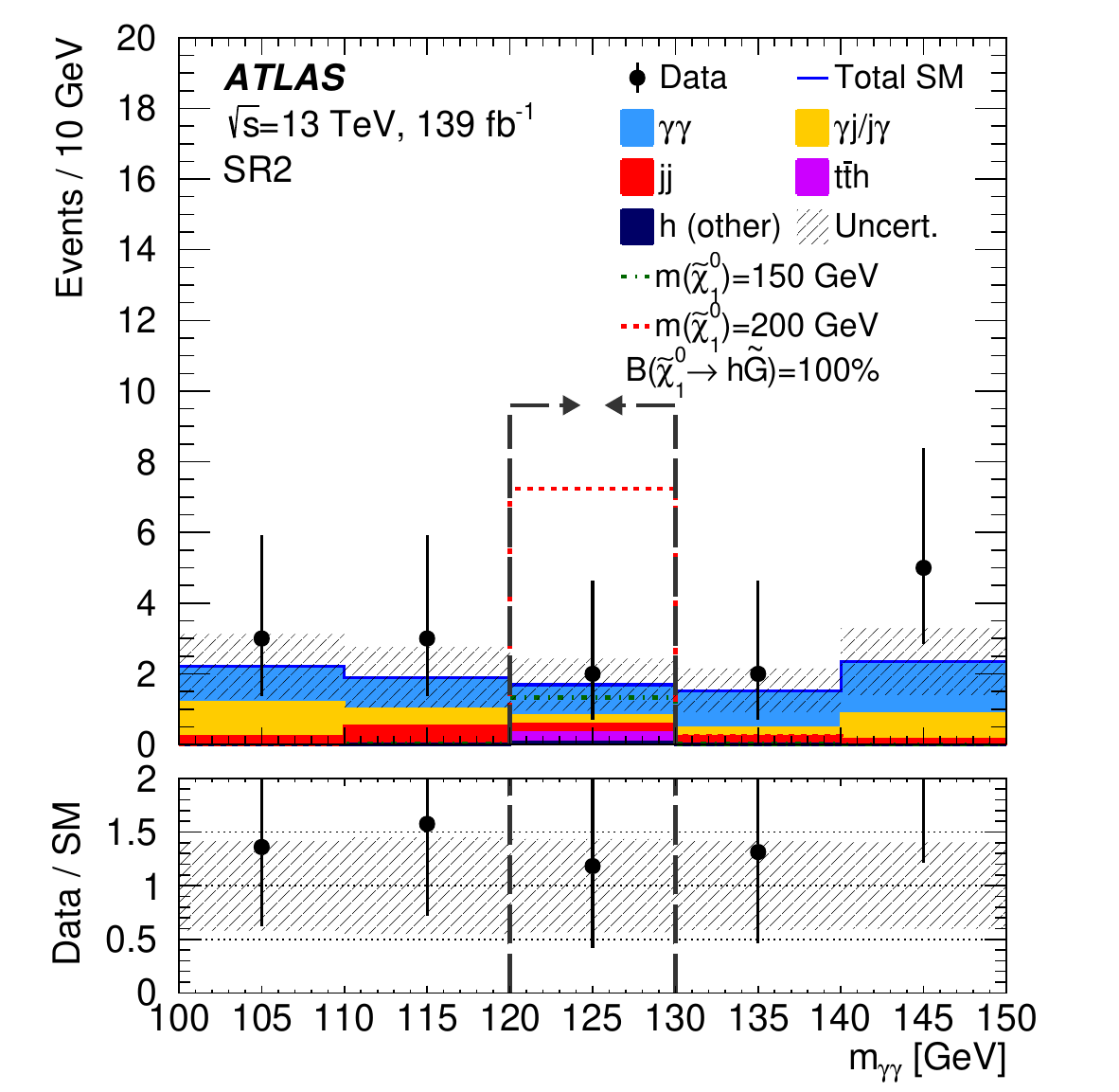}
\caption[]{}
\end{subfigure}
\caption{\small Distribution of the diphoton invariant mass with all selections of the signal regions applied, except on $\myy$ itself, for the three signal regions: (a) SR1h, (b) SR1Z and (c) SR2. The background estimation techniques described in the text are applied. The different backgrounds are stacked to add up to the total SM prediction in each bin. The predicted yields for signal benchmark models of varying \ninoone{} mass are also overlaid (not stacked) assuming $\BFNh$ to equal (a) $100\%$, (b) $50\%$ and (c) $100\%$. Background and signal predictions are normalised to the luminosity. The background category `$h$ (other)' includes events originating from VBF, \VH, ggF, \tHq, \tHW and \bbH, all subdominant in this signature. The sizes of the statistical and systematic uncertainties are indicated by the shaded areas. The lower panels show the ratio of the data to the SM prediction. Arrows indicate the borders of the signal region ($|\myy-125\,\GeV|<5\,\GeV$).
The first and last bins include the underflows and overflows respectively.
}
\label{fig:srplots}
\end{figure*}

The $\myy$ distributions associated with the three signal regions are
shown in Figure~\ref{fig:srplots}.
The observed yields in each SR, reported in Table~\ref{tab:results}, are found to be consistent with the corresponding SM predictions,
and inconsistent with substantial additional contributions from the supersymmetric signal models under test.
Table~\ref{tab:results} also shows, for each SR, the upper limits at 95\% confidence level (CL) on the possible number of beyond the Standard Model (BSM) events $S^{95}$, and on the visible BSM cross-section  $\langle\epsilon{ \sigma}\rangle_{\textrm{obs}}^{95} = S_{\textrm{obs}}^{95}/\int\!{\cal L}\textrm{d}t$, defined as the product of the ATLAS experimental acceptance and efficiency and the production cross-section.
 
\begin{table}[t!b]
\centering
\setlength{\tabcolsep}{0.0pc}
 
\caption{Observed and expected numbers of events in the three signal regions. The background category `$h$ (other)' includes events originating from VBF, \VH, ggF, \tHq, \tHW and \bbH, all subdominant in this signature. The table also includes
model-independent 95\% CL upper limits on the visible number of BSM events ($S^{95}_{\textrm{obs}}$), the number of BSM events given the expected number of background events ($S^{95}_{\textrm{exp}}$) and the visible BSM cross-section ($\langle\epsilon{ \sigma}\rangle_{\textrm{obs}}^{95}$), all calculated from pseudo-experiments. The discovery $p$-value ($p_0$) is also shown and its value is capped at 0.5 if the observed number of events is below the expected number of events.}
\begin{tabular*}{ 0.8\textwidth}{@{\extracolsep{\fill}}lccc}
\toprule
\textbf{Channel}           & \textbf{SR1h}            & \textbf{SR1Z}            & \textbf{SR2}              \\
\midrule
Observed           & 3              & 5              & 2                    \\
\midrule
Total SM               & $3.9 \pm 0.6$          & $6.4 \pm 1.0$          & $1.7 \pm 0.7$              \\
\midrule
$\gamma\gamma$           & $2.5 \pm 0.5$          & $3.7 \pm 0.7$ & $0.88 \pm 0.26$              \\
$\gamma j$          & $0.47 \pm 0.28$          & $0.8 \pm 0.5$ & $0.24 \pm 0.15$              \\
$j\gamma$          &  $0.088 \pm 0.014$          & $0.27 \pm 0.04$          & $0.00 \pm 0.6$              \\
$jj$          &  $ < 0.01 $         & $0.07 \pm 0.05$          & $0.22_{-0.22}^{+0.24}$              \\
\ttH          & $0.41 \pm 0.04$          & $0.297 \pm 0.025$          & $0.27 \pm 0.06$              \\
$h$ (other)       & $0.40 \pm 0.08$          & $1.22 \pm 0.26$          & $0.064 \pm 0.011$              \\
\midrule
 
$\langle\epsilon{ \sigma}\rangle_{\textrm{obs}}^{95}$ [fb]  &  $0.03$ &   $0.04$ &  $0.03$ \\
 
$S_{\textrm{obs}}^{95}$  & $4.8$ & $5.5$ &  $4.8$ \\

$S_{\textrm{exp}}^{95}$ &  $ { 5.4 }^{ +2.2 }_{ -1.5 }$ &  $ { 6.7 }^{ +2.6 }_{ -1.8 }$ &  $ { 4.6 }^{ +1.6 }_{ -0.8 }$ \\

$p_0$ & $ 0.50$ & $ 0.50$ & $ 0.43$ \\
 
\bottomrule
\end{tabular*}

\label{tab:results}
\end{table}
 
\begin{figure*}[p!]
\centering
\includegraphics[width=.7\textwidth]{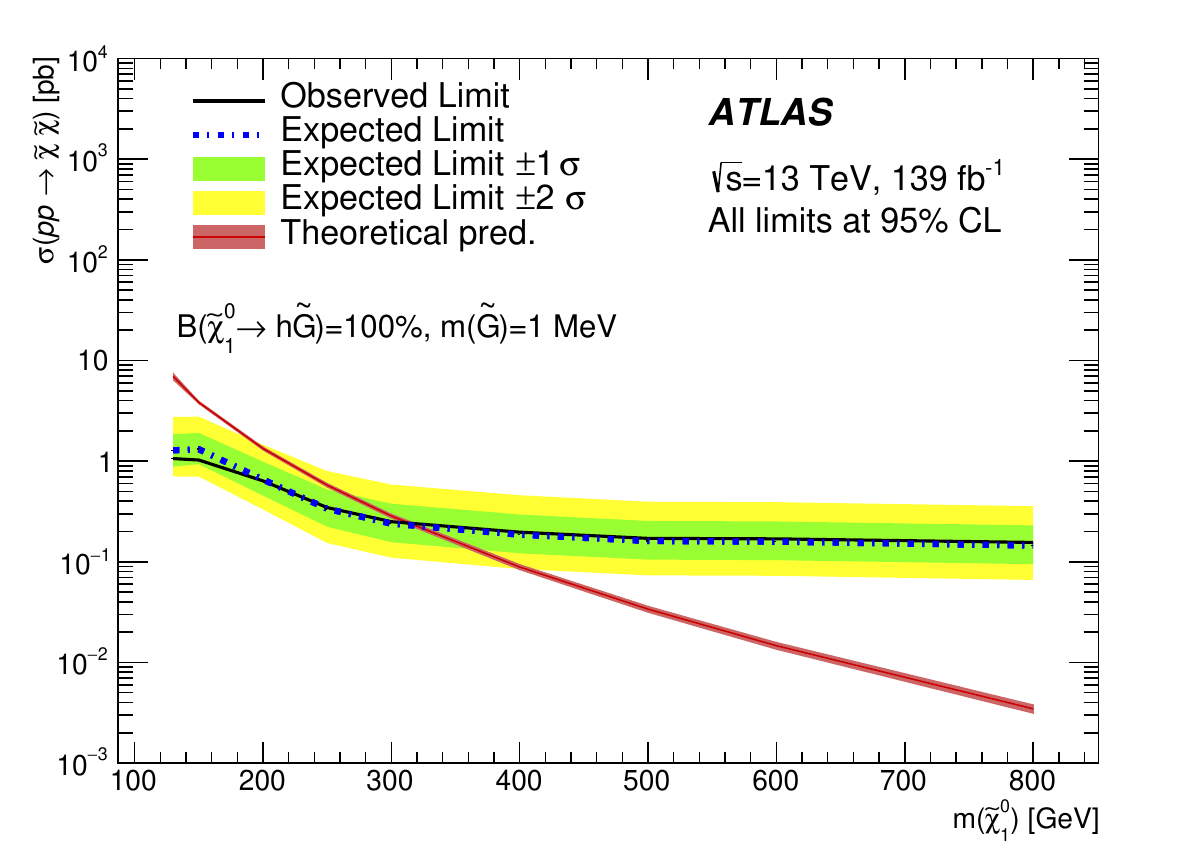}
\caption{\small Observed and expected limits on the pure higgsino cross-section at 95\% CL assuming $\BFNh=100\%$ for different $\ninoone$ masses, obtained by a statistical combination of the three signal regions SR1h, SR1Z and SR2. The inner and outer bands indicate the $1\sigma$ and $2\sigma$ variation on the expected limit respectively.}
\label{fig:exclusion_yybb}
\end{figure*}
 
\begin{figure*}[p!]
\centering
\includegraphics[width=.7\textwidth]{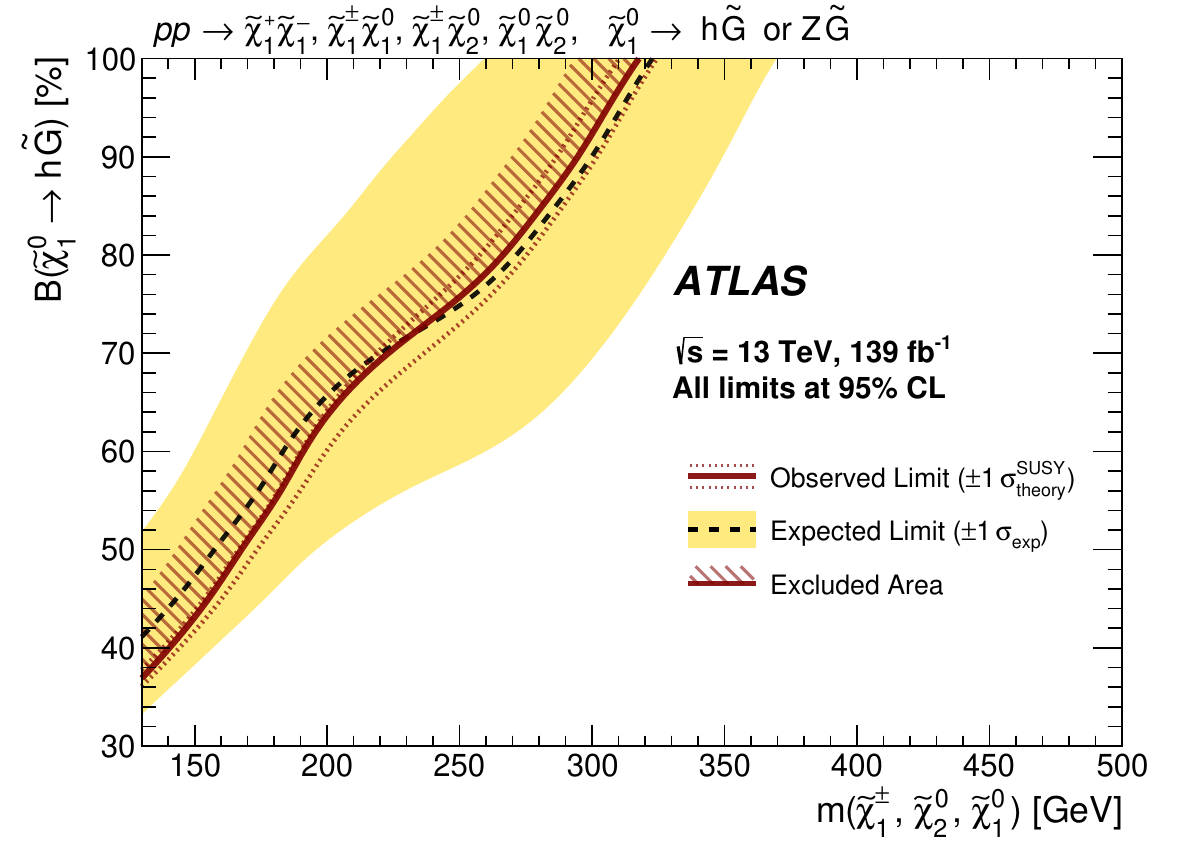}
\caption{\small Observed and expected 95\% CL limits on the pure-higgsino branching fraction to $\BFNh$ as a function of the higgsino mass $m(\ninoone)$ assuming it decays via either $\ninoone\rightarrow h\tilde{G}$ or $\ninoone\rightarrow Z\tilde{G}$. Limits are obtained by performing a statistical combination of the three signal regions SR1h, SR1Z and SR2. The $\pm\,1\sigma$ variation on the expected limit is shown. The dotted lines indicate the observed limit obtained by a variation of theoretical prediction for the neutralino production cross-section by $\pm1\,\sigma$. Values of $\BFNh$ larger than the observed 95\% CL limit are excluded, as indicated by the hatched area.}
\label{fig:exclusion_2D_yybb}
\end{figure*}
 
A statistical combination of the three signal regions is performed, and expected and observed limits on the cross-section for higgsino pair-production are again calculated following the $CL_{\text{s}}$ formalism~\cite{Read:2002hq} implemented in HistFitter~\cite{Baak:2014wma}, while using the profile likelihood ratio as a test statistic~\cite{Cowan:2010js}.
The results of this statistical combination are shown in Figure~\ref{fig:exclusion_yybb} for the assumption $\BFNh=100\%$.
Under the same assumption, cross-sections above $1\,\mathrm{pb}$ are excluded at 95\%~CL for $m(\ninoone) < 150\,\GeV$, and the theoretical prediction for the pure higgsino cross-section is excluded at 95\%~CL for neutralino masses $ < 320\,\GeV$.
 
The branching fraction assumption can be relaxed in such a way that the higgsino is allowed to decay either via $\ninoone \rightarrow h \gravino$ or via $\ninoone \rightarrow Z\gravino$ (assuming $\BFNh+\BFNZ =1$). Limits for this scenario are set as a function of higgsino mass and branching fraction, as shown in Figure~\ref{fig:exclusion_2D_yybb}.
As expected, the sensitivity is strongest in the region with smaller higgsino mass,
and larger $\BFNh$.
Nevertheless, the exclusion at 95\% CL extends down to $\BFNh=36\%$ for neutralino masses of $130\,\GeV$.
The sensitivity in the low-mass regime ($m(\ninoone) \lesssim 200\,\GeV$) is improved over the search in final states with multiple b-jets presented in Ref.~\cite{ATLAS:2024tqe}.
 
\FloatBarrier
 
In conclusion, a search is performed with the ATLAS detector using 139 fb$^{-1}$
of proton--proton collisions at a centre-of-mass energy of $\ensuremath{\sqrt{s} =\SI{13}{\TeV}}$ for excess events containing either pairs of Higgs bosons,
or one Higgs boson with one $\Zboson$ boson, in a final state containing a photon pair, a \(b\bar{b}\) pair and missing transverse momentum.
The observed data is in agreement with the SM prediction in all three tested signal regions and
the results are used to derive upper limits on the pair production of higgsinos in gauge-mediated supersymmetry models.
The observed 95\% CL exclusion is extended up to
$m(\ninoone) = 320\,\GeV$ for $\BFNh=100\%$
and down to $\BFNh=36\%$ for $m(\ninoone)=130\,\GeV$.
This analysis provides complementarity to previous ATLAS searches in final states with multiple leptons or multiple $b$-jets, targeting different decays of the electroweak bosons.
 
\section*{Acknowledgements}


We thank CERN for the very successful operation of the LHC and its injectors, as well as the support staff at
CERN and at our institutions worldwide without whom ATLAS could not be operated efficiently.
 
The crucial computing support from all WLCG partners is acknowledged gratefully, in particular from CERN, the ATLAS Tier-1 facilities at TRIUMF/SFU (Canada), NDGF (Denmark, Norway, Sweden), CC-IN2P3 (France), KIT/GridKA (Germany), INFN-CNAF (Italy), NL-T1 (Netherlands), PIC (Spain), RAL (UK) and BNL (USA), the Tier-2 facilities worldwide and large non-WLCG resource providers. Major contributors of computing resources are listed in Ref.~\cite{ATL-SOFT-PUB-2023-001}.
 
We gratefully acknowledge the support of ANPCyT, Argentina; YerPhI, Armenia; ARC, Australia; BMWFW and FWF, Austria; ANAS, Azerbaijan; CNPq and FAPESP, Brazil; NSERC, NRC and CFI, Canada; CERN; ANID, Chile; CAS, MOST and NSFC, China; Minciencias, Colombia; MEYS CR, Czech Republic; DNRF and DNSRC, Denmark; IN2P3-CNRS and CEA-DRF/IRFU, France; SRNSFG, Georgia; BMBF, HGF and MPG, Germany; GSRI, Greece; RGC and Hong Kong SAR, China; ISF and Benoziyo Center, Israel; INFN, Italy; MEXT and JSPS, Japan; CNRST, Morocco; NWO, Netherlands; RCN, Norway; MEiN, Poland; FCT, Portugal; MNE/IFA, Romania; MESTD, Serbia; MSSR, Slovakia; ARIS and MVZI, Slovenia; DSI/NRF, South Africa; MICINN, Spain; SRC and Wallenberg Foundation, Sweden; SERI, SNSF and Cantons of Bern and Geneva, Switzerland; NSTC, Taipei; TENMAK, T\"urkiye; STFC/UKRI, United Kingdom; DOE and NSF, United States of America.
 
Individual groups and members have received support from BCKDF, CANARIE, CRC and DRAC, Canada; CERN-CZ, PRIMUS 21/SCI/017 and UNCE SCI/013, Czech Republic; COST, ERC, ERDF, Horizon 2020, ICSC-NextGenerationEU and Marie Sk{\l}odowska-Curie Actions, European Union; Investissements d'Avenir Labex, Investissements d'Avenir Idex and ANR, France; DFG and AvH Foundation, Germany; Herakleitos, Thales and Aristeia programmes co-financed by EU-ESF and the Greek NSRF, Greece; BSF-NSF and MINERVA, Israel; Norwegian Financial Mechanism 2014-2021, Norway; NCN and NAWA, Poland; La Caixa Banking Foundation, CERCA Programme Generalitat de Catalunya and PROMETEO and GenT Programmes Generalitat Valenciana, Spain; G\"{o}ran Gustafssons Stiftelse, Sweden; The Royal Society and Leverhulme Trust, United Kingdom.
 
In addition, individual members wish to acknowledge support from Chile: Agencia Nacional de Investigaci\'on y Desarrollo (FONDECYT 1190886, FONDECYT 1210400); China: National Natural Science Foundation of China (NSFC - 12175119, NSFC 12275265, NSFC-12075060); Czech Republic: PRIMUS Research Programme (PRIMUS/21/SCI/017); EU: H2020 European Research Council (ERC - 101002463, H2020-MSCA-IF-2020: HPOFHIC - 10103); European Union: European Research Council (ERC - 948254), Horizon 2020 Framework Programme (MUCCA - CHIST-ERA-19-XAI-00), European Union, Future Artificial Intelligence Research (FAIR-NextGenerationEU PE00000013), Italian Center for High Performance Computing, Big Data and Quantum Computing (ICSC, NextGenerationEU), Marie Sklodowska-Curie Actions (EU H2020 MSC IF GRANT NO 101033496); France: Agence Nationale de la Recherche (ANR-20-CE31-0013, ANR-21-CE31-0013, ANR-21-CE31-0022), Investissements d'Avenir Idex (ANR-11-LABX-0012), Investissements d'Avenir Labex (ANR-11-LABX-0012); Germany: Baden-Württemberg Stiftung (BW Stiftung-Postdoc Eliteprogramme), Deutsche Forschungsgemeinschaft (DFG - CR 312/5-1); Italy: Istituto Nazionale di Fisica Nucleare (FELLINI G.A. n. 754496, ICSC, NextGenerationEU); Japan: Japan Society for the Promotion of Science (JSPS KAKENHI JP21H05085, JSPS KAKENHI JP22H01227, JSPS KAKENHI JP22H04944); Netherlands: Netherlands Organisation for Scientific Research (NWO Veni 2020 - VI.Veni.202.179); Norway: Research Council of Norway (RCN-314472); Poland: Polish National Agency for Academic Exchange (PPN/PPO/2020/1/00002/U/00001), Polish National Science Centre (NCN 2021/42/E/ST2/00350, NCN UMO-2019/34/E/ST2/00393, UMO-2020/37/B/ST2/01043, UMO-2021/40/C/ST2/00187); Slovenia: Slovenian Research Agency (ARIS grant J1-3010); Spain: BBVA Foundation (LEO22-1-603), Generalitat Valenciana (Artemisa, FEDER, IDIFEDER/2018/048), La Caixa Banking Foundation (LCF/BQ/PI20/11760025), Ministry of Science and Innovation (MCIN \& NextGenEU PCI2022-135018-2, MICIN \& FEDER PID2021-125273NB, RYC2019-028510-I, RYC2020-030254-I, RYC2021-031273-I, RYC2022-038164-I), PROMETEO and GenT Programmes Generalitat Valenciana (CIDEGENT/2019/023, CIDEGENT/2019/027); Sweden: Swedish Research Council (VR 2018-00482, VR 2022-03845, VR 2022-04683, VR grant 2021-03651), Knut and Alice Wallenberg Foundation (KAW 2017.0100, KAW 2018.0157, KAW 2018.0458, KAW 2019.0447); Switzerland: Swiss National Science Foundation (SNSF - PCEFP2\_194658); United Kingdom: Leverhulme Trust (Leverhulme Trust RPG-2020-004); United States of America: U.S. Department of Energy (ECA DE-AC02-76SF00515), Neubauer Family Foundation.


\printbibliography
 
\clearpage
\input{atlas_authlist}

\end{document}

%% file: atlas_authlist.tex
 
\begin{flushleft}
\hypersetup{urlcolor=black}
{\Large The ATLAS Collaboration}

\bigskip

\AtlasOrcid[0000-0002-6665-4934]{G.~Aad}$^\textrm{\scriptsize 103}$,
\AtlasOrcid[0000-0002-5888-2734]{B.~Abbott}$^\textrm{\scriptsize 121}$,
\AtlasOrcid[0000-0002-1002-1652]{K.~Abeling}$^\textrm{\scriptsize 55}$,
\AtlasOrcid[0000-0001-5763-2760]{N.J.~Abicht}$^\textrm{\scriptsize 49}$,
\AtlasOrcid[0000-0002-8496-9294]{S.H.~Abidi}$^\textrm{\scriptsize 29}$,
\AtlasOrcid[0000-0002-9987-2292]{A.~Aboulhorma}$^\textrm{\scriptsize 35e}$,
\AtlasOrcid[0000-0001-5329-6640]{H.~Abramowicz}$^\textrm{\scriptsize 152}$,
\AtlasOrcid[0000-0002-1599-2896]{H.~Abreu}$^\textrm{\scriptsize 151}$,
\AtlasOrcid[0000-0003-0403-3697]{Y.~Abulaiti}$^\textrm{\scriptsize 118}$,
\AtlasOrcid[0000-0003-0762-7204]{A.C.~Abusleme~Hoffman}$^\textrm{\scriptsize 138a}$
% ...
\bigskip
\\

% ...
$^{29}$Physics Department, Brookhaven National Laboratory, Upton NY; United States of America.\\
% ...
$^{35}$$^{(a)}$Facult\'e des Sciences Ain Chock, Universit\'e Hassan II de Casablanca;$^{(b)}$Facult\'{e} des Sciences, Universit\'{e} Ibn-Tofail, K\'{e}nitra;$^{(c)}$Facult\'e des Sciences Semlalia, Universit\'e Cadi Ayyad, LPHEA-Marrakech;$^{(d)}$LPMR, Facult\'e des Sciences, Universit\'e Mohamed Premier, Oujda;$^{(e)}$Facult\'e des sciences, Universit\'e Mohammed V, Rabat;$^{(f)}$Institute of Applied Physics, Mohammed VI Polytechnic University, Ben Guerir; Morocco.\\
% ...
$^{49}$Fakult\"{a}t Physik , Technische Universit{\"a}t Dortmund, Dortmund; Germany.\\
% ...
$^{55}$II. Physikalisches Institut, Georg-August-Universit\"{a}t G\"ottingen, G\"ottingen; Germany.\\
% ...
$^{103}$CPPM, Aix-Marseille Universit\'e, CNRS/IN2P3, Marseille; France.\\
% ...
$^{118}$Department of Physics, New York University, New York NY; United States of America.\\
% ...
$^{121}$Homer L. Dodge Department of Physics and Astronomy, University of Oklahoma, Norman OK; United States of America.\\
% ...
$^{138}$$^{(a)}$Departamento de F\'isica, Pontificia Universidad Cat\'olica de Chile, Santiago;$^{(b)}$Millennium Institute for Subatomic physics at high energy frontier (SAPHIR), Santiago;$^{(c)}$Instituto de Investigaci\'on Multidisciplinario en Ciencia y Tecnolog\'ia, y Departamento de F\'isica, Universidad de La Serena;$^{(d)}$Universidad Andres Bello, Department of Physics, Santiago;$^{(e)}$Instituto de Alta Investigaci\'on, Universidad de Tarapac\'a, Arica;$^{(f)}$Departamento de F\'isica, Universidad T\'ecnica Federico Santa Mar\'ia, Valpara\'iso; Chile.\\
% ...
$^{151}$Department of Physics, Technion, Israel Institute of Technology, Haifa; Israel.\\
$^{152}$Raymond and Beverly Sackler School of Physics and Astronomy, Tel Aviv University, Tel Aviv; Israel.\\
% ...
$^{a}$ Also Affiliated with an institute covered by a cooperation agreement with CERN.\\
$^{b}$ Also at An-Najah National University, Nablus; Palestine.\\
$^{c}$ Also at Borough of Manhattan Community College, City University of New York, New York NY; United States of America.\\
$^{d}$ Also at Center for High Energy Physics, Peking University; China.\\
$^{e}$ Also at Center for Interdisciplinary Research and Innovation (CIRI-AUTH), Thessaloniki; Greece.\\
$^{f}$ Also at Centro Studi e Ricerche Enrico Fermi; Italy.\\
% ...

\end{flushleft}
